\documentclass[aps,preprintnumbers,groupedaddress]{revtex4}%
\usepackage{amsmath}
\usepackage{graphicx}
\usepackage{amsfonts}
\usepackage{amssymb}%
\providecommand{\U}[1]{\protect\rule{.1in}{.1in}}

\begin{document}
\preprint{arXiv:0811.2417}
\title{Flavor Hierarchy From F-theory}
\author{Jonathan J. Heckman}
\email{jheckman@fas.harvard.edu}
\author{Cumrun Vafa}
\email{vafa@physics.harvard.edu}
\affiliation{Jefferson Physical Laboratory, Harvard University, Cambridge, MA 02138, USA}

\begin{abstract}
\noindent{It has recently been shown that F-theory based constructions provide
a potentially promising avenue for engineering GUT models which descend to the
MSSM. In this note we show that in the presence of background fluxes, these models automatically achieve hierarchical Yukawa matrices in the quark and lepton sectors. At leading order, the existence of a $U(1)$ symmetry which is related
to phase rotations of the internal holomorphic coordinates at the brane
intersection point leads to rank one Yukawa matrices. Subleading corrections
to the internal wave functions from variations in the background fluxes generate small violations of this $U(1)$, leading to hierarchical
Yukawa structures reminiscent of the Froggatt-Nielsen mechanism. The expansion
parameter for this perturbation is in terms of $\sqrt{\alpha_{GUT}}$.
Moreover, we naturally obtain a hierarchical CKM matrix with $V_{12} \sim
V_{21} \sim\varepsilon$, $V_{23} \sim V_{32} \sim\varepsilon^{2}$, $V_{13}
\sim V_{31} \sim\varepsilon^{3}$, where $\varepsilon\sim\sqrt{\alpha_{GUT}}$,
in excellent agreement with observation.}

\end{abstract}
\maketitle

\section{Introduction}

Flavor physics remains a poorly understood aspect of string based
constructions as well as phenomenological models which aim to reproduce the
Standard Model or MSSM at low energies. In this note we show that the minimal
form of the F-theory GUT models recently developed in \cite{BHVI,BHVII,HVWEAK}
(see also
\cite{DonagiWijnholtI,TatarHETF,IbanezSOFT,ConlonWAVE,MARSANOSWEET,DonagiWijnholtII,FontIbanez}%
) \textit{automatically} generates viable Yukawa matrices in the quark and
lepton sectors. In the present context, a minimal implementation of an
F-theory GUT simply reflects the discrete choice in the geometry that we
consider models with the fewest possible number of matter curves and
intersection points necessary for compatibility with the MSSM.

The MSSM superpotential contains the terms:
\begin{equation}
W_{MSSM}=\lambda_{u}^{ij}\cdot Q^{i}U^{j}H_{u}+\lambda_{d}^{ij}\cdot
Q^{i}D^{j}H_{d}+\lambda_{l}^{ij}\cdot L^{i}E^{j}H_{d}+\cdots
\end{equation}
where $i$ and $j$ are generation indices so that $U^{3}$ refers to the
right-handed top quark, and the $\lambda$'s denote $3\times3$ matrices.
Although genericity arguments might suggest all of the associated masses
should be comparable, there is a well-known hierarchy \cite{PDG}:%
\begin{align}
\left(  m_{u},m_{c},m_{t}\right)   &  \sim(0.003,1.3,170)\times\text{GeV
}\label{up}\\
\left(  m_{d},m_{s},m_{b}\right)   &  \sim(0.005,0.1,4)\times\text{GeV
\ \ \ }\label{down}\\
\left(  m_{e},m_{\mu},m_{\tau}\right)   &  \sim(0.0005,0.1,1.8)\times
\text{GeV.} \label{lep}%
\end{align}
To leading order, the inter-generational ratios of masses between particles
with the same gauge quantum numbers are insensitive to running effects. On the
other hand, running to the GUT\ scale reduces the heaviest generation quark
masses by roughly a factor of three.

Letting $V_{u,d}^{L,R}$ denote unitary matrices such that $V_{u}^{L}%
\lambda_{u}V_{u}^{R\dag}$ and $V_{d}^{L}\lambda_{d}V_{d}^{R\dag}$ are
diagonal, the norm of elements in the CKM\ matrix $V_{CKM}\equiv V_{u}%
^{L}V_{d}^{L\dag}$ are also hierarchical \cite{PDG}:%
\begin{equation}
\left\vert V_{CKM}(M_{weak})\right\vert \sim\left(
\begin{array}
[c]{ccc}%
0.97 & 0.23 & 0.004\\
0.23 & 0.97 & 0.04\\
0.008 & 0.04 & 0.99
\end{array}
\right)  \text{.} \label{CKMPDG}%
\end{equation}
This is encapsulated in the Wolfenstein parameterization of $V_{CKM}$
\cite{WolfParam}:%
\begin{equation}
V_{CKM}=\left(
\begin{array}
[c]{ccc}%
1-\lambda^{2}/2 & \lambda & A\lambda^{3}\left(  \rho-i\eta\right) \\
-\lambda & 1-\lambda^{2}/2 & A\lambda^{2}\\
A\lambda^{3}\left(  1-\rho-i\eta\right)  & -A\lambda^{2} & 1
\end{array}
\right)  +O(\lambda^{4})
\end{equation}
where $\lambda\sim0.23$, $A\sim0.82$, $\rho\sim0.22$, $\eta\sim0.34$ at the
weak scale. To one loop order, only $A$ evolves with scale
\cite{PokorskiWolf,RamondStitch}. In fact, even including two loop running
effects, $A$ only changes by an order one factor \cite{RossSerna}.

This type of hierarchy is potentially present in models with a global $U(1)$
flavor symmetry as in the Froggatt-Nielsen mechanism \cite{FroggattNielsen}.
For example, assigning FN\ charges $4,2,0$ to the three generations of $Q$,
$U$ superfields and introducing a chiral superfield $X_{FN}$ with FN\ charge
$-1$, the superpotential term $\left(  X_{FN}/M_{FN}\right)  ^{2(3-i)+2(3-j)}%
\cdot Q^{i}U^{j}$ will induce hierarchical Yukawa couplings once $X_{FN}$
develops a suitable vev. Similar considerations apply for the remaining Yukawa matrices.

On the other hand, the three generations all possess the same gauge quantum
numbers with respect to the Standard Model gauge group. For this reason,
introducing an horizontal symmetry may appear somewhat ad hoc. In this note we
show that an approximate symmetry of the local geometry plays a similar role
to that present in the Froggatt-Nielsen mechanism. Nevertheless, we also find
that the exact structure of the Yukawa matrices differs from single field
Froggatt-Nielsen models due to the profiles of the internal wave functions.

\section{Partially Twisted Gauge Theory and F-theory GUTs}

In F-theory GUT\ models (see \cite{HVLHC,WijnholtREV} for reviews), the gauge
fields of the MSSM\ descend from the eight-dimensional worldvolume of a
seven-brane wrapping a complex del Pezzo surface $S$ with a GUT gauge group.
This GUT group is broken to $SU(3)_{C}\times SU(2)_{L}\times U(1)_{Y}$ via an
internal flux through the seven-brane in the $U(1)_{Y}$ direction of the GUT
group \cite{BHVII,DonagiWijnholtII}. The chiral matter of the MSSM descends
from zero modes of six-dimensional fields localized on matter curves in $S$.
In the classical limit, the zero mode wave functions vanish off of the matter
curve. Strictly speaking, this is only approximately true because these wave
functions have non-vanishing support in directions transverse to the matter
curve. In purely gauge theoretic terms, these wave functions can be derived
from an appropriate internal local Higgsing of a parent eight-dimensional
theory \cite{KV}. In this parent theory, there is an adjoint-valued field
$\phi$. When $\phi$ develops a vev which breaks the parent theory gauge group,
adjoint-valued fermions in the parent eight-dimensional gauge multiplet
descend to \textquotedblleft bifundamentals\textquotedblright\ which are
trapped along complex codimension one matter curves defined by the vanishing
locus $\phi=0$, much as in the Nielsen-Olesen vortex \cite{NielsenOlesen}. The
trapped zero modes obey the system of equations \cite{BHVI}:%
\begin{align}
\omega_{S}\wedge\partial_{A}\psi_{\alpha}+\frac{i}{2}\left[  \overline{\phi
},\chi_{\alpha}\right]   &  =0 + O\left(  \frac{M_{GUT}}{M_{\ast}}\right) \label{gaugecond}\\
\overline{\partial}_{A}\chi_{\alpha}-\left[  \phi,\psi_{\alpha}\right]   &  =0 + O\left(  \frac{M_{GUT}}{M_{\ast}}\right)
\label{chiequation}%
\end{align}
where $\omega_{S}$ denotes the K\"{a}hler form on $S$, and $\chi_{\alpha}$ as
well as $\psi_{\alpha}$ denote fermions in the 8d gauge multiplet. The
subscript $A$ reflects the background internal gauge field from the GUT
seven-brane, as well as potentially other seven-branes which intersect the GUT
seven-brane. In addition the ``$O\left(  \frac{M_{GUT}}{M_{\ast}}\right)$'' reflect the possibility of contributions
from higher dimension operators induced by other background fluxes of the compactification associated with more general
p-form potentials. Here, $M_{\ast}$ is the characteristic mass scale of the F-theory compactification which is related to
the GUT scale $M_{GUT}$ and $\alpha_{GUT}$ as \cite{BHVII}:%
\begin{equation}
M_{\ast}^{4}=\alpha_{GUT}^{-1}\cdot M_{GUT}^{4}\text{.}%
\end{equation}
In general, the effects of these background fluxes can non-trivially mix with such higher-form potentials.
For example, in an abelian gauge theory on a D-brane, the presence of an NS B-field shifts the
field strength as:
\begin{equation}\label{mixer}
F^{\prime} = F + B.
\end{equation}

For simplicity of presentation, consider the case where the
background field strength is parallel to the matter curve. Letting $z_{\bot}$
denote the coordinate normal to the curve $\phi=0$, and $z$ the coordinate on
the curve $\phi=0$, the trapped zero mode wave function is of the form:%
\begin{equation}
\Psi\sim f(z,\overline{z})\exp(-\gamma z_{\bot}\overline{z_{\bot}})\text{,}
\label{GaussLead}%
\end{equation}
where $f$ is a $\overline{\partial}_{A}$ zero mode of the bundle defined by
the gauge field along the curve $\phi=0$ , and $\gamma$ depends on the overall
scale specified by the K\"{a}hler form. Note that the ratio of any two zero
modes of $\overline{\partial}_{A}$ can be expressed purely in terms of
holomorphic functions of $z$, a fact that will be crucial in the analysis to
follow. Further note that by rescaling the K\"{a}hler form so that
$\gamma\rightarrow\infty$, the choice of representative used in the
topological field theory can be made arbitrarily peaked along the matter curve.

The intersection of three matter curves in $S$ leads to Yukawa couplings among
the corresponding matter fields. The Yukawa couplings between the Higgs fields
and chiral matter are given by triple overlap integrals of the form:%
\begin{equation}
\lambda^{ij}=\underset{S}{\int}\Lambda\Psi^{i}\Phi^{j}\text{.} \label{overlap}%
\end{equation}
Here, $\Lambda\Psi^{i}\Phi^{j}$ may be viewed as shorthand for the triple
product of the wave functions $\Lambda_{H_{u}}\Psi_{Q}^{i}\Psi_{U}^{j}$,
$\Lambda_{H_{d}}\Psi_{Q}^{i}\Psi_{D}^{j}$ or $\Lambda_{H_{d}}\Psi_{L}^{i}%
\Psi_{E}^{j}$ in the obvious notation. In the limit where the internal field
variations (other than the Higgs) is constant, this integral reduces to a
product of the three wave functions at the various intersection points of the
matter curves:%
\begin{equation}
\lambda^{ij}=\underset{p}{\sum}\Lambda(p)\Psi^{i}(p)\Phi^{j}(p)
\label{yukclass}%
\end{equation}
where $p$ denotes a point of triple intersection. Indeed, this superpotential
is the descendant of the superpotential for bulk modes in the parent theory
which makes no reference to the K\"{a}hler form. As such, we are free to
rescale the K\"{a}hler form dependence until the overlap of the corresponding
wave functions is only non-zero at the point of mutual intersection.

This is quite analogous to the computation of perturbative
Yukawa couplings in the context of $(2,2)$ sigma models, where in the
classical limit the Yukawa couplings are given by classical intersection
theory. In the A-model, this result is then deformed by worldsheet instanton
corrections and in the B-model by the variation of the complex structure. The
setup in F-theory is parallel to the B-model setup and we can view the above
Yukawa coupling as the analog of the limit of \textquotedblleft
large\textquotedblright\ complex structure. Just as in the B-model at finite
complex structure, there will also be subleading corrections to equation
(\ref{yukclass}). Such corrections can occur as a result of higher dimension
operator contributions to the effective action, which will deform
equations (\ref{gaugecond}) and (\ref{chiequation}).

From a bottom up perspective, a minimal implementation of the chiral matter of
an $SU(5)$ GUT requires one matter curve for the $10_{M}$ and one for the
$\overline{5}_{M}$ where $SU(5)$ respectively enhances to $SO(10)$ and
$SU(6)$. The Higgs $5_{H}$ and $\overline{5}_{H}$ localize on distinct curves
where $SU(5)$ enhances to $SU(6)$. The background flux through these curves
dictates the number of chiral matter zero modes in the low energy theory,
while a suitable background flux through the Higgs curves achieves
doublet-triplet splitting \cite{BHVII}.

The superpotential terms $5_{H}\times10_{M}\times10_{M}$ and $\overline{5}%
_{H}\times\overline{5}_{M}\times10_{M}$ respectively originate from
enhancements to $E_{6}$ and $SO(12)$. Geometrically, this requires the
$10_{M}$ curve to intersect itself, as well as the $5_{H}$ curve at a single
point, and for the $10_{M}$ curve to intersect the $\overline{5}_{M}$ and
$\overline{5}_{H}$ curves at another point of $S$ \cite{BHVII}. We shall
respectively label these triple intersection points as $p_{up}$ and $p_{down}%
$. While additional intersections are in principle possible, the minimal, and
most generic situation will typically only contain one intersection point of
each type. Note that in principle there can be additional points where the
singularity type enhances since the matter curve may consist of more than
one irreducible component. In the special case where the wave
functions are given identically by a sharply peaked Gaussian which is
concentrated along a matter curve, a single triple intersection for
each type of Yukawa leads to rank one Yukawa
matrices \cite{BHVII}. For example, the up quark Yukawas are given by:%
\begin{equation}
\lambda_{u}^{ij}=\Lambda_{H_{u}}(p_{up})\Psi_{Q}^{i}(p_{up})\Psi_{U}%
^{j}(p_{up})
\end{equation}
which is manifestly a rank one matrix. If there were no further corrections,
the resulting theory would contain a massive top quark, and massless up and
charm quarks. Similar considerations hold for the down type and lepton masses.

In evaluating subleading corrections to the Yukawas, it is more natural to go
to a canonical orthonormal basis of wave functions where this rank one
structure in the Yukawa matrix is more manifest. To this end, let $z_{1}$ and
$z_{2}$ denote two local coordinates on $S$ such that $z_{1}=0$ and $z_{2}=0$
denote the matter curves where the wave functions $\Phi^{j}$ and$\ \Psi^{i}$
of equation (\ref{overlap}) respectively localize. By an appropriate unitary
change of basis, we can always arrange for $\Psi^{i}\sim z_{1}^{3-i}$ and
$\Phi^{j}\sim z_{2}^{3-j}$ along the respective curves where the wave
functions localize (in particular we can choose these wave functions to be
orthonormal by allowing subleading pieces in the powers of $z_{i}$ in each
wave function). In this basis, $\lambda^{33}\neq0$ whereas all the other
components of $\lambda^{ij}$ vanish. In terms of the wave function overlap
integral of equation (\ref{overlap}) this vanishing structure in the Yukawa
matrix can be ascribed to the presence of a geometric $U(1)_{1}\times
U(1)_{2}$ action on the $z_{j}$:%
\begin{equation}
z_{j}\rightarrow\exp(i\alpha_{j})z_{j}\text{.}%
\end{equation}
The terms of the integrand which vanish are precisely those which are not
invariant under this $U(1)_{1}\times U(1)_{2}$ symmetry \cite{footnoteONE}.

\section{Wave Function Distortion and Yukawa Hierarchies}

We now compute subleading corrections in the Yukawa matrices due to
distortions in the profile of the internal wave functions. In principle,
such distortions from various fluxes induced by p-form potential of the
compactification. The ones of relevance for distorting the Yukawas correspond to higher dimension
operator deformations of equations (\ref{gaugecond}) and (\ref{chiequation}). The overlap
integral of interest is given by:%
\begin{equation}
\lambda^{ij}={\int}d^{2}z_{1}d^{2}z_{2}\cdot\Omega\left(  z_{1},\overline
{z_{1}},z_{2},\overline{z_{2}}\right)  \cdot\left(  \frac{z_{1}}{R_{1}%
}\right)  ^{3-i}\left(  \frac{z_{2}}{R_{2}}\right)  ^{3-j} \label{lamint}%
\end{equation}
where $R_{1}$ and $R_{2}$ denote the characteristic lengths of the curves
$z_{1}=0$ and $z_{2}=0$, and $\Omega\left(  z_{1},\overline{z_{1}}%
,z_{2},\overline{z_{2}}\right)  $ is a function which includes the
contributions from the\ Higgs field wave function, as well as other
non-holomorphic terms associated with the other wave functions. Note that all
information about flavor is contained in the term $z_{1}^{3-i}z_{2}^{3-j}$.

As noted above, a non-zero contribution to $\lambda^{ij}$ can only occur when
a contribution to the integrand is invariant under the local $U(1)_{1}\times
U(1)_{2}$ action of the coordinates. A non-zero entry for the $\lambda^{ij}$
when $i$ and $j$ are both different from $3$ therefore requires some
contribution from $\Omega$ to carry non-trivial $U(1)_{1}\times U(1)_{2}$ charge.

Such corrections originate from general fluxes of the compactification. Since $\phi$ is
holomorphic and in the local geometry lies in a $U(1)\times U(1)$ subgroup of the parent local gauge
symmetry, without loss of generality we can choose local coordinates so that
$\phi_{1}=z_{1}$ and $\phi_{2}=z_{2}$ in the neighborhood of the intersection
point. Our aim will be to characterize possible violations of the
$U(1)_{1} \times U(1)_{2}$ symmetry due to background fluxes. Owing to the symmetries of equations (\ref{gaugecond}) and (\ref{chiequation}) and their deformation by possible higher dimension operators, the leading order
distortion of the wave functions from background fluxes is a linear
combination of the $U(1)_{1}\times U(1)_{2}$ invariants $z_{i}\overline{z_{j}}$ so that:%
\begin{equation}
\Psi=\Psi^{(0)}\cdot\exp\left(  \mathcal{M}^{i\overline{j}}z_{i}%
\overline{z_{j}}\right)  \label{ourPSI}%
\end{equation}
where $\Psi^{(0)}$ denotes the contribution to the wave function in the limit
where the field strength vanishes, and $\mathcal{M}^{i\overline{j}}$ is a
$2\times2$ matrix. Although the off-diagonal contributions to this quadratic form indeed violate
the \textquotedblleft axial\textquotedblright\ combination of $U(1)_{1}$ and
$U(1)_{2}$, such terms cannot contribute to the Yukawa couplings
\cite{axialNOTE}.

We stress that the distortion due to $\mathcal{M}$ can in principle originate not just from gauge fields,
but from more general fluxes induced by p-form potentials of the compactification, which will also mix with these gauge fields.
Indeed, while gauge field flux alone does not turn out to distort the Yukawa couplings, other fluxes do \cite{CCHV}. Further note
that by dimensional analysis, the overall scaling of $\mathcal{M}$ is fixed as $\mathcal{M} \sim M_{GUT}^{2}$.

More generally, the field strengths will vary over points of the geometry. In
an adiabatic approximation we can simply retain the form of $\Psi$ but with
$\mathcal{M}^{i\overline{j}}$ now given by a function of $z_{1},\overline
{z_{1}},z_{2}$ and $\overline{z_{2}}$. We now perform a series expansion in
derivatives of $\mathcal{M}^{i\overline{j}}$ around the point of triple
intersection:%
\begin{equation}
\mathcal{M}^{i\overline{j}}z_{i}\overline{z_{j}}=\underset{k,l,m,n}{\sum}%
\frac{\partial_{1}^{k}\overline{\partial}_{\overline{1}}^{l}\partial_{2}%
^{m}\overline{\partial}_{\overline{2}}^{n}\mathcal{M}^{i\overline{j}}%
(0)}{k!l!m!n!}\cdot z_{1}^{k}\overline{z_{1}}^{l}z_{2}^{m}\overline{z_{2}}%
^{n}\cdot z_{i}\overline{z_{j}}\text{.}%
\end{equation}
The crucial point is that the monomial $z_{1}^{k}\overline{z_{1}}^{l}z_{2}%
^{m}\overline{z_{2}}^{n}$ has $U(1)_{1}\times U(1)_{2}$ charge $\left(
k-l,m-n\right)  $. Expanding the exponential of equation (\ref{ourPSI}), it
now follows that there will generically be corrections to the rank one Yukawa coupling.

We now compute the leading order behavior of the Yukawa matrix entry
$\lambda^{ij}$ by performing a series expansion of $\Omega$ in equation
(\ref{lamint}). $U(1)_{1}\times U(1)_{2}$ charge conservation implies that the
only terms from a series expansion of $\Omega$ which can contribute are of the
form:%
\begin{equation}
\left(  z_{1}\overline{z_{1}}\right)  ^{a}\left(  z_{2}\overline{z_{2}%
}\right)  ^{b}\cdot(\overline{z_{1}})^{3-i}(\overline{z_{2}})^{3-j}\text{.}
\label{abexpan}%
\end{equation}
Each power of $z$ reflects the presence of a derivative in the series expansion.

There are in principal two expansion parameters available which will lead to
different hierarchies in the Yukawa couplings. The first, and perhaps most
obvious possibility is that we can simply count the number of total $z$'s
which appear in a given term. This amounts to performing a derivative
expansion in the gauge field strength. In this case, the leading order
contributions will always originate from terms where $a=b=0$. The contribution
from the derivative expansion is given as:%
\begin{equation}
\delta\lambda_{DER}^{ij}=c^{ij}\cdot\frac{\overline{\partial}_{\overline{1}%
}^{3-i}\overline{\partial}_{\overline{2}}^{3-j}\mathcal{M}(0)}{M_{\ast
}^{8-i-j}}\cdot\left(  \frac{1}{R_{1}M_{\ast}}\right)  ^{3-i}\left(  \frac
{1}{R_{2}M_{\ast}}\right)  ^{3-j} \label{DEREXP}%
\end{equation}
where $\mathcal{M}$ is shorthand for the generic entry $\mathcal{M}%
^{i\overline{j}}$, $c^{ij}$ is an order one coefficient which reflects the
evaluation of the integral in equation (\ref{lamint}).

On the other hand, there is another natural expansion parameter given by
expanding in successive powers of the first gradient of the flux. The leading
order perturbation to the Yukawas from this class of terms is:%
\begin{equation}
\delta\lambda_{FLX}^{ij}=c^{ij}\cdot\left(  \frac{\overline{\partial
}_{\overline{1}}\mathcal{M}(0)}{R_{1}M_{\ast}^{4}}\right)  ^{3-i}\left(
\frac{\overline{\partial}_{\overline{2}}\mathcal{M}(0)}{R_{2}M_{\ast}^{4}%
}\right)  ^{3-j}\text{.}%
\end{equation}

To estimate the magnitude of each $\delta\lambda^{ij}$, we note by dimensional analysis, $\mathcal{M} \sim M_{GUT}^2$.
Further, each successive derivative of
$\mathcal{M}$ contributes an extra factor of $1/R_{1}$ or $1/R_{2}$. Setting:%
\begin{equation}
\varepsilon_{0}=\mathcal{M}(0)/M_{\ast}^{2}\text{, }\varepsilon_{1}=\left(
M_{\ast}R_{1}\right)  ^{-2}\text{, }\varepsilon_{2}=\left(  M_{\ast}%
R_{2}\right)  ^{-2}\text{,}%
\end{equation}
we thus obtain:%
\begin{align}
\delta\lambda_{DER}^{ij}  &  =c^{ij}\cdot\varepsilon_{0}\cdot\left(
\varepsilon_{1}\right)  ^{3-i}\left(  \varepsilon_{2}\right)  ^{3-j}\\
\delta\lambda_{FLX}^{ij}  &  =c^{ij}\cdot\left(  \varepsilon_{0}%
\varepsilon_{1}\right)  ^{3-i}\left(  \varepsilon_{0}\varepsilon_{2}\right)
^{3-j}%
\end{align}
adopting a parametrization where $\varepsilon_{1}\sim\varepsilon_{2}\sim
\kappa$ and $\varepsilon^{2}\sim\varepsilon_{0}\varepsilon_{2}$, the Yukawa
coupling matrix therefore assumes the form:%
\begin{equation}
\lambda=\lambda_{(0)}+\delta\lambda_{DER}+\delta\lambda_{FLX}\sim\left(
\begin{array}
[c]{ccc}%
0 & 0 & 0\\
0 & 0 & 0\\
0 & 0 & 1
\end{array}
\right)  +\left(
\begin{array}
[c]{ccc}%
\kappa^{3}\cdot\varepsilon^{2} & \kappa^{2}\cdot\varepsilon^{2} & \kappa
\cdot\varepsilon^{2}\\
\kappa^{2}\cdot\varepsilon^{2} & \kappa\cdot\varepsilon^{2} & \varepsilon
^{2}\\
\kappa\cdot\varepsilon^{2} & \varepsilon^{2} & \varepsilon_{0}%
\end{array}
\right)  +\left(
\begin{array}
[c]{ccc}%
\varepsilon^{8} & \varepsilon^{6} & \varepsilon^{4}\\
\varepsilon^{6} & \varepsilon^{4} & \varepsilon^{2}\\
\varepsilon^{4} & \varepsilon^{2} & 1
\end{array}
\right)
\end{equation}
where each entry is multiplied by an order one constant. At this point, it is
important to note that although the analysis crucially depends on the
violation of a $U(1)$ symmetry, the structure of $\delta\lambda_{DER}$ is
somewhat different from the simplest Froggatt-Nielsen parametrization
mentioned earlier. Note, however, that $\delta\lambda_{FLX}$ is indeed
consistent with a single field Froggatt-Nielsen model.

In the geometry, there are in fact three $\kappa$'s and three $\varepsilon$'s
because there are three distinct Yukawa matrices of interest. Indeed, the
background hyperflux will generate order one distortions in the lepton doublet
and down quark wave functions so that the entries of $\lambda_{l}$ and
$\lambda_{d}$ can differ by order one constants, so that there is no a priori
mass relation between the down type quarks and the charged leptons
\cite{BHVII}.

Because $R_{i}\sim M_{GUT}^{-1}$, one might at first think that since
$\varepsilon\sim\kappa\sim\sqrt{\alpha_{GUT}}$, that the $DER$ expansion will
always dominate over the $FLX$ expansion. However, there can be a further
enhancement in the $FLX$ expansion because it is more strongly on how chiral matter couples to
background fluxes. Since $\sqrt{\alpha_{GUT}}$ is not
extremely small, the $FLX$ expansion can in principle dominate over the $DER$
expansion in Yukawas involving fields which have large couplings to the background flux.

Determining whether $\delta\lambda_{DER}$ or $\delta\lambda_{FLX}$ dominates
at each order depends on how the left and right chiral matter wave functions
couple to the background fluxes. In F-theory GUTs, a background flux in the hypercharge
direction is always present \cite{BHVII,DonagiWijnholtII}, so to illustrate this point,
consider the coupling of the matter wave functions to this flux. We shall assume that
the strength of this coupling to background fluxes holds for other more general fluxes since in general, these higher form fluxes
can mix with the gauge fields, as in equation (\ref{mixer}) so we shall use it as a rough guide to determine which type of expansion dominates for a given type of particle species. In the case of hyperflux, the internal wave functions couple to this particular background flux in strength
proportional to their hypercharges. It follows that the hyperflux contribution to $\mathcal{M}%
^{i\overline{j}}$ is proportional to the corresponding hypercharge. The term
$\delta\lambda_{FLX}^{ij}$ can dominate over $\delta\lambda_{DER}^{ij}$
provided:%
\begin{equation}
\left(  Y_{\max}\cdot\frac{\overline{\partial}F(0)}{RM_{\ast}^{4}}\right)
^{2}>Y_{\max}\cdot\frac{\overline{\partial}^{2}F(0)}{M_{\ast}^{4}}\cdot\left(
\frac{1}{RM_{\ast}}\right)  ^{2} \label{qestimate}%
\end{equation}
where $F$ denotes the generic field strength, and $Y_{\max}$ denotes in an
integral normalization the maximal norms of the hypercharge for the left and
right chiral matter wave functions associated to each Yukawa. In particular,
we have:
\begin{equation}%
\begin{tabular}
[c]{|c|c|c|c|}\hline
& lepton & up & down\\\hline
$Y_{\max}$ & $6$ & $4$ & $2$\\\hline
\end{tabular}
\ \ \text{.}%
\end{equation}
Using the representative scaling $F/M_{\ast}^{2}\sim1/\left(  RM_{\ast
}\right)  ^{2}\sim\alpha_{GUT}^{1/2}$ yields:%
\begin{equation}
\mathcal{F}\equiv Y_{\max}\cdot\alpha_{GUT}^{1/2}\gtrsim1,
\end{equation}
where $\mathcal{F}$ measures the relative strengths of the flux and gradient
expansions. Plugging in $\alpha_{GUT}^{1/2}\sim0.2$ as well as the integral
values for the hypercharges yields:%
\begin{equation}%
\begin{tabular}
[c]{|c|c|c|c|}\hline
& lepton & up & down\\\hline
$\mathcal{F}$ & $1.2\sim1$ & $0.8\sim1$ & $0.4<1$\\\hline
\end{tabular}
\ \ \text{.}%
\end{equation}
This rough computation suggests that for the down type Yukawas, $\delta
\lambda_{DER}$ dominates, whereas for the up type and charged lepton Yukawas,
$\delta\lambda_{FLX}^{ij}$ could potentially dominate. Even though this is by
itself quite heuristic, we will shortly find that comparison with observation
indeed corroborates this picture.

Setting $\kappa\sim\varepsilon$ the Yukawa matrices for $\lambda_{DER}$ and
$\lambda_{FLX}$ are:%
\begin{equation}
\lambda_{DER}\sim\left(
\begin{array}
[c]{ccc}%
\varepsilon^{5} & \varepsilon^{4} & \varepsilon^{3}\\
\varepsilon^{4} & \varepsilon^{3} & \varepsilon^{2}\\
\varepsilon^{3} & \varepsilon^{2} & 1
\end{array}
\right)  \text{, }\lambda_{FLX}\sim\left(
\begin{array}
[c]{ccc}%
\varepsilon^{8} & \varepsilon^{6} & \varepsilon^{4}\\
\varepsilon^{6} & \varepsilon^{4} & \varepsilon^{2}\\
\varepsilon^{4} & \varepsilon^{2} & 1
\end{array}
\right)
\end{equation}
this leads to distinct mass hierarchies:%
\begin{align}
\lambda_{DER}  &  \Rightarrow m_{1}:m_{2}:m_{3}\sim\varepsilon^{5}%
:\varepsilon^{3}:1\\
\lambda_{FLX}  &  \Rightarrow m_{1}:m_{2}:m_{3}\sim\varepsilon^{8}%
:\varepsilon^{4}:1\text{.}%
\end{align}

Next consider diagonalization of $\lambda$ via the matrices $V^{L}$ and
$V^{R}$. In fact, it is enough to diagonalize $\lambda\lambda^{\dag}$:%
\begin{equation}
\left(  \lambda_{DER}\right)  \left(  \lambda_{DER}\right)  ^{\dag}\sim\left(
\begin{array}
[c]{ccc}%
\varepsilon^{6} & \varepsilon^{5} & \varepsilon^{3}\\
\varepsilon^{5} & \varepsilon^{4} & \varepsilon^{2}\\
\varepsilon^{3} & \varepsilon^{2} & 1
\end{array}
\right)  \text{, }\left(  \lambda_{FLX}\right)  \left(  \lambda_{FLX}\right)
^{\dag}\sim\left(
\begin{array}
[c]{ccc}%
\varepsilon^{8} & \varepsilon^{6} & \varepsilon^{4}\\
\varepsilon^{6} & \varepsilon^{4} & \varepsilon^{2}\\
\varepsilon^{4} & \varepsilon^{2} & 1
\end{array}
\right)  \text{.}%
\end{equation}
Note that these matrices have a structure of the form $\varepsilon
^{a_{i}+a_{j}}$ which is consistent with a single field Froggatt-Nielsen model
with charges $3,2,0$ for $\left(  \lambda_{DER}\right)  \left(  \lambda
_{DER}\right)  ^{\dag}$ and $4,2,0$ for $\left(  \lambda_{FLX}\right)  \left(
\lambda_{FLX}\right)  ^{\dag}$. In this case, the matrices $V^{L}$ and $V^{R}$
are of the form $\varepsilon^{\left\vert a_{i}-a_{j}\right\vert }$ so that:%
\begin{equation}
V_{DER}^{L}\sim V_{DER}^{R}\sim\left(
\begin{array}
[c]{ccc}%
1 & \varepsilon & \varepsilon^{3}\\
\varepsilon & 1 & \varepsilon^{2}\\
\varepsilon^{3} & \varepsilon^{2} & 1
\end{array}
\right)  \text{, }V_{FLX}^{L}\sim V_{FLX}^{R}\sim\left(
\begin{array}
[c]{ccc}%
1 & \varepsilon^{2} & \varepsilon^{4}\\
\varepsilon^{2} & 1 & \varepsilon^{2}\\
\varepsilon^{4} & \varepsilon^{2} & 1
\end{array}
\right)  \text{.}%
\end{equation}

In the above we have focussed on the overlap of the wave functions at a single
point of the geometry. At each intersection point, there is a local choice of
$U(1)$'s with respect to which the given mass hierarchy of the generations is
manifest. Indeed, this is a basis independent statement.

On the other hand, the CKM matrix $V_{CKM}=V_{u}^{L}V_{d}^{L\dag}$ measures
the mismatch between the matrices which diagonalize the Yukawas in the
\textit{same} basis. Preserving the hierarchy of the CKM matrix requires that
the hierarchy present in $V_{u}^{L}$ and $V_{d}^{L}$ is manifest in the same
basis for the $Q$ wave functions. For this to be the case, the eigenspace
decomposition defined by the local $U(1)$'s at the two different interaction
points must remain relatively unchanged. This occurs provided the distance
between the intersection points $p_{up}$ and $p_{down}$ (both of which lie on
the $Q$ matter curve) should be less than $0.1\times M_{GUT}^{-1}$. Note that
this is to be compared with the length scale of the curves, which is roughly
$M_{GUT}^{-1}$. While this is certainly possible, and involves only a very
mild fine tuning in the location of the two points of triple intersection, it
is suggestive of a higher unification structure which could naturally
accommodate the strict identification $p_{up}=p_{down}$. The presence of a
higher unification structure is also in line with the appearance of a
$U(1)_{PQ}$ factor in F-theory GUT\ scenarios which incorporate the effects of
supersymmetry breaking \cite{HVWEAK}. For the purposes of this paper, we shall
only assume that $p_{up}$ and $p_{down}$ are sufficiently close that this
subtlety can be neglected.

Under this mild assumption, it is in fact now possible to determine the form
of the CKM matrix. When either $V^{L}$ matrix is determined by the derivative
expansion, $\kappa\sim\varepsilon$ dominates over order $\varepsilon^{2}$
terms so that:
\begin{equation}
V_{CKM}^{F-th}\left(  \varepsilon\right)  \sim\left(
\begin{array}
[c]{ccc}%
1 & \varepsilon & \varepsilon^{3}\\
\varepsilon & 1 & \varepsilon^{2}\\
\varepsilon^{3} & \varepsilon^{2} & 1
\end{array}
\right)  \sim\left(
\begin{array}
[c]{ccc}%
1 & \alpha_{GUT}^{1/2} & \alpha_{GUT}^{3/2}\\
\alpha_{GUT}^{1/2} & 1 & \alpha_{GUT}\\
\alpha_{GUT}^{3/2} & \alpha_{GUT} & 1
\end{array}
\right)  \text{.} \label{initioCKM}%
\end{equation}
By inspection, the $3,2,0$ Froggatt-Nielsen charges dominate in the
CKM\ matrix. In addition, the gap in the Froggatt-Nielsen charges can be
traced back to the tensor structure of the flux, which distinguishes it from a scalar.
In the less likely possibility (which is indeed not realized) where both
$V^{L}$'s are given by $V_{FLX}^{L}$, the form of $V_{CKM}$ is again of the
form $V_{FLX}^{L}$.

\section{Comparison with Experiment}

In the previous section we obtained a parametrization of the up, down and
lepton Yukawas in terms of the parameters $\varepsilon$ and $\kappa$. In this
section, we show that the most natural estimates for $\varepsilon$ and
$\kappa$ dictated by the GUT\ structure are in beautiful accord with
observation. Using the explicit values:%
\begin{equation}
\kappa\sim\varepsilon\sim M_{GUT}^{2}/M_{\ast}^{2}\sim\alpha_{GUT}^{1/2}%
\sim0.2\text{,}%
\end{equation}
we can now estimate the CKM\ matrix to be:%
\begin{equation}
V_{CKM}^{F-th}\sim\left(
\begin{array}
[c]{ccc}%
1 & \alpha_{GUT}^{1/2} & \alpha_{GUT}^{3/2}\\
\alpha_{GUT}^{1/2} & 1 & \alpha_{GUT}\\
\alpha_{GUT}^{3/2} & \alpha_{GUT} & 1
\end{array}
\right)  \sim\left(
\begin{array}
[c]{ccc}%
1 & 0.2 & 0.008\\
0.2 & 1 & 0.04\\
0.008 & 0.04 & 1
\end{array}
\right)  \text{.}%
\end{equation}
Comparing with the measured values \cite{PDG}:
\begin{equation}
\left\vert V_{CKM}(M_{weak})\right\vert \sim\left(
\begin{array}
[c]{ccc}%
0.97 & 0.23 & 0.004\\
0.23 & 0.97 & 0.04\\
0.008 & 0.04 & 0.99
\end{array}
\right)  \text{,}%
\end{equation}
reveals a beautiful match between theory and observation! Even though we
should only have expected an agreement in parameters up to order one effects,
we find this close match with the data reassuring.

We also find close agreement with the expected mass hierarchies in the quark
and charged lepton sectors. Due to the different structure of the
perturbations $\delta\lambda_{DER}$ and $\delta\lambda_{FLX}$, there are in
principle two different hierarchical mass ratios given by:%
\begin{align}
\lambda_{DER}  &  \Rightarrow m_{1}:m_{2}:m_{3}\sim\varepsilon^{5}%
:\varepsilon^{3}:1\label{DERRAT}\\
\lambda_{FLX}  &  \Rightarrow m_{1}:m_{2}:m_{3}\sim\varepsilon^{8}%
:\varepsilon^{4}:1\text{.} \label{FLXRAT}%
\end{align}

As discussed previously, the dominant contribution to $\lambda_{d}$ is
expected to be from $\delta\lambda_{FLX}$. On the other hand, we have observed
that the dominant contribution to $\lambda_{u}$ and $\lambda_{l}$ could in
principle be either $\delta\lambda_{DER}$ or $\delta\lambda_{FLX}$.

To compare with observation, we can fit the observed masses to a given value
of $\varepsilon$. At the present crude level of analysis, it is enough to
neglect the effects of running in such a match because we will be taking third
and fourth roots of order one numbers. It turns out that the best match to our
scenario is achieved when both $\lambda_{u}$ and $\lambda_{l}$ are of the
latter type with corresponding values for $\varepsilon$:
\begin{equation}%
\begin{tabular}
[c]{|l|l|l|l|}\hline
& $l_{FLX}$ & $u_{FLX}$ & $d_{DER}$\\\hline
$\varepsilon$ & $0.36$ & $0.26$ & $0.27$\\\hline
\end{tabular}
\ \label{tabtab}%
\end{equation}
which are all within order one factors of $\alpha_{GUT}^{1/2}\sim0.2$! Fixing
the values of the top, bottom and tau mass to their observed values, we can
now use the ratios of (\ref{DERRAT}) and (\ref{FLXRAT}) to extract the masses
of the lighter generations. Comparing the F-theory result with the observed
values yields:%
\begin{align}
\left(  m_{u},m_{u}^{F-th}\right)  ,\left(  m_{c},m_{c}^{F-th}\right)
,\left(  m_{t},m_{t}^{F-th}\right)   &  \sim
(0.003,0.004),(1.3,0.8),(170,170)\times\text{GeV }\\
\left(  m_{d},m_{d}^{F-th}\right)  ,\left(  m_{s},m_{s}^{F-th}\right)
,\left(  m_{b},m_{b}^{F-th}\right)   &  \sim
(0.005,0.006),(0.1,0.08),(4,4)\times\text{GeV \ \ \ }\\
\left(  m_{e},m_{e}^{F-th}\right)  ,\left(  m_{\mu},m_{\mu}^{F-th}\right)
,\left(  m_{\tau},m_{\tau}^{F-th}\right)   &  \sim
(0.0005,0.0005),(0.1,0.03),(1.8,1.8)\times\text{GeV,}%
\end{align}
which by inspection, are quite similar.

The parametrization of the up and charged lepton Yukawas in terms of
$\delta\lambda_{DER}$ does not reliably reproduce these same mass ratios. One
way to see this is to restore the explicit $\varepsilon$ and $\kappa$
dependence in the mass ratios. There are two independent ratios of masses, so
in this case we can exactly solve for $\varepsilon$ and $\kappa$ with the
result:
\begin{equation}%
\begin{tabular}
[c]{|l|l|l|l|}\hline
& $l_{DER}$ & $u_{DER}$ & $d_{DER}$\\\hline
$\varepsilon$ & $0.88$ & $0.40$ & $0.33$\\\hline
$\kappa$ & $0.07$ & $0.05$ & $0.22$\\\hline
\end{tabular}
\text{ \ .}%
\end{equation}
Note that the $\varepsilon$'s for $l_{DER}$ and $u_{DER}$ are both greater
than $\alpha_{GUT}^{1/2}$, whereas the $\kappa$'s are both smaller. Indeed,
the ratio $\varepsilon/\kappa$ for these two cases is now an order ten number,
rather than the order one parameter expected from general considerations.
Note, however, that the ratio $\varepsilon/\kappa$ for $d_{DER}$ is indeed an
order one number, which is in accord with our scenario.

The match to the CKM matrix and the mass parametrization of table
(\ref{tabtab}) lends considerable credence to the simple picture of wave
function distortion we have found. We note that although we have implicitly
worked within the framework of the MSSM, the hierarchies we have found are
more general, and only require a supersymmetric structure near the GUT scale.

Up to now, we have only discussed the hierarchy in the mass ratios. In fact,
our model also predicts the actual masses for the top, bottom and tau mass as
well \cite{BHVII}:%
\begin{align}
m_{t}^{F-th}(M_{GUT})  &  \sim\alpha_{GUT}^{3/4}\cdot\langle H_{u}%
\rangle=\alpha_{GUT}^{3/4}\cdot\left\langle H\right\rangle \sin\beta\\
m_{b}^{F-th}(M_{GUT})  &  \sim m_{\tau}^{F-th}(M_{GUT})=\alpha_{GUT}%
^{3/4}\cdot\langle H_{d}\rangle=\alpha_{GUT}^{3/4}\cdot\left\langle
H\right\rangle \cos\beta\text{.}%
\end{align}
Using the value $\left\langle H\right\rangle \sim170\text{ GeV}$, and assuming
the large value $\tan\beta\sim30$ common to the models discussed in
\cite{HVWEAK} yields:%
\begin{align}
m_{t}^{F-th}(M_{GUT})  &  \sim20\text{ GeV}\\
m_{b}^{F-th}(M_{GUT})  &  \sim m_{\tau}^{F-th}(M_{GUT})\sim0.6\text{ GeV,}%
\end{align}
which are to be compared with the observed values run up to the GUT scale.
These running effects are roughly given by a factor of three reduction for the
quark masses so that:%
\begin{align}
m_{t}(M_{GUT})  &  \sim55\text{ GeV}\\
m_{b}(M_{GUT})  &  \sim m_{\tau}(M_{GUT})\sim1.5\text{ GeV}%
\end{align}
which matches the values of our scenario up to order one factors (in fact, a
factor of three). In tandem with our estimate for the mass ratios, we thus
obtain a crude estimate for the masses of all the quarks and charged leptons
in terms of $\left\langle H\right\rangle $, $\tan\beta$ and $\alpha_{GUT}$.

\section{Discussion}

In this note we have shown that simply demanding a local F-theory GUT\ model
with the most generic features automatically implies hierarchical structure in
the Yukawa couplings. Moreover, we have found that simple estimates on the
behavior of the wave functions near points of triple intersection are in
remarkable accord with observation! While it is certainly quite satisfying to
see this type of structure naturally emerge from a string based model, it is
also possible to go further.

The first point is that the general form of the matrices we have found can in
principle be extended to an arbitrary number of generations. Indeed, it is
interesting to ask whether this model constrains the existence of additional
generations of quarks \cite{SCHWCONV}. For example, in a four generation
model, the analogue of equation (\ref{initioCKM}) is now determined by the FN
charges $4,3,2,0$, leading to:%
\begin{equation}
V_{CKM}^{(4-gen)}\sim\left(
\begin{array}
[c]{cccc}%
1 & \varepsilon & \varepsilon^{2} & \varepsilon^{4}\\
\varepsilon & 1 & \varepsilon & \varepsilon^{3}\\
\varepsilon^{2} & \varepsilon & 1 & \varepsilon^{2}\\
\varepsilon^{4} & \varepsilon^{3} & \varepsilon^{2} & 1
\end{array}
\right)  \text{.}%
\end{equation}
Mixing in the three lightest generations corresponds to the upper left
$3\times3$ block of this matrix, which is in worse agreement with experiment
than the three generation model. Thus, we conclude that independent of other
experimental constraints, simply obtaining the observed CKM\ matrix structure
for the three lightest generations in our model \textit{predicts} that there
is no fourth generation of quarks!

We have also seen that the generic profile of the internal wave functions
generates a rank three Yukawa matrix. In particular, this implies that the up
quark is not massless, in accord with results from lattice gauge theory
simulations of QCD. In particular, this implies that the strong CP\ problem is
real. Interestingly, supersymmetry breaking in F-theory GUT\ models
automatically contain an axion with a phenomenologically viable value for the
decay constant \cite{BHVII}.

We find it quite remarkable that the most minimal geometric structures in this
framework are in fact sufficient for the purposes of achieving
phenomenologically viable hierarchies in the Yukawa couplings. Turning the
analysis around, the beautiful match obtained in minimal F-theory GUT
scenarios can be viewed as an important constraint on the class of
compactifications, and in particular the requisite geometries which can
reproduce detailed features of the MSSM.

Finally, one can extend this analysis to the case of neutrino masses and
mixing. Whereas the quarks and charged leptons localize on matter curves in
$S$, in some F-theory GUT\ models the right-handed neutrinos $N_{R}$ localize
on matter curves which only touch $S$ at a point \cite{BHVII}. We have argued
that the Yukawa matrices for the quarks and charged leptons can exhibit
similar hierarchies provided the two points of intersection $p_{up}$ and
$p_{down}$ are sufficiently close together. Assuming appropriate wave function
overlaps so that the superpotential terms $LN_{R}H_{u}$ and $M_{maj}N_{R}%
N_{R}$ are present will generate viable masses for the neutrinos via the
seesaw mechanism. Note, moreover, that this type of geometry can also easily
accommodate large mixing angles provided the interaction point leading to
$LEH_{d}$ on the $L$-matter curve is not finely tuned to be near the
interaction point leading to $LN_{R}H_{u}$. We are currently working out the
details of this scenario.

\emph{Acknowledgements} The work of the authors is supported in part by NSF
grant PHY-0244821.

\section{Appendix: Wave Function Distortion From Fluxes}

In this Appendix we provide additional details on the profile of wave
functions in the presence of a background gauge field strength. We refer the
interested reader to \cite{BHVII} for a more detailed analysis of wave
function distortion effects due to gauge field strengths in the direction
parallel to the matter curve. We present this example to illustrate how wave function
distortion works in general. Indeed, it turns out that though gauge field fluxes alone do not distort the
profile of the Yukawas, these contributions can mix with higher form potentials, leading to an appropriate
Yukawa distortion \cite{CCHV}.

We first consider the case where this gauge field strength is constant, and
then explain how the profile of the wave function changes in the presence of
variations in this field. To be completely explicit, we consider the case of a
bulk $E_{6}$ theory with $27$'s localized along three matter curves of rank
one enhancement to $E_{7}$ which form a triple intersection at a point where
$E_{6}$ enhances to $E_{8}$. In the parent $E_{8}$ theory, the background
values of the scalar $\phi$ and internal gauge fields satisfy the
BPS\ equations of motion \cite{BHVI}:%
\begin{equation}
\overline{\partial}_{A}\phi=0\text{, }F^{(2,0)}=0\text{, }F^{(0,2)}=0\text{,
}\omega_{S}\wedge F^{(1,1)}=0\text{.}%
\end{equation}
The $E_{6}$ theory is defined by Higgsing the parent $E_{8}$ theory in the
direction: $\left\langle \phi\right\rangle =\left\langle \phi_{1}\right\rangle
T_{1}+\left\langle \phi_{2}\right\rangle T_{2}$ where the $T_{i}$ denote two
Cartan generators in the $SU(3)$ factor of $E_{6}\times SU(3)\subset E_{8}$.
The loci of partial enhancement to $E_{7}$ are then given by $\left\langle
\phi_{1}\right\rangle =0$, $\left\langle \phi_{2}\right\rangle =0$ and
$\left\langle \phi_{1}\right\rangle +\left\langle \phi_{2}\right\rangle =0$.
In addition, we also consider a background gauge field strength taking values
in the Cartan of $SU(3)$.

Adopting a local system of coordinates $z_{1}$, $z_{2}$ such that
$z_{i}=\left\langle \phi_{i}\right\rangle $, the modes trapped along the curve
$z_{1}=0$ curve satisfy equations (\ref{gaugecond}) and (\ref{chiequation}):%
\begin{align}
v_{2}\left(  \partial_{1}+A_{1}\right)  \psi_{\overline{1}}+v_{1}\left(
\partial_{2}+A_{2}\right)  \psi_{\overline{2}}+\overline{z_{1}}\chi_{12}  &
=0\label{one}\\
\left(  \overline{\partial}_{\overline{1}}+A_{\overline{1}}\right)  \chi
_{12}+z_{1}\psi_{\overline{1}}  &  =0\label{two}\\
\left(  \overline{\partial}_{\overline{2}}+A_{\overline{2}}\right)  \chi
_{12}+z_{1}\psi_{\overline{2}}  &  =0 \label{three}%
\end{align}
where in the above, we have taken a canonical presentation for the K\"{a}hler
form. Similar equations hold for the other massless modes of the theory. When
the background field strength is constant, the solutions to this system of
equations depend to leading order on the fluxes through a term of the form
$\exp(\mathcal{M}^{i\overline{j}}z_{i}\overline{z_{j}})$, where $\mathcal{M}%
^{i\overline{j}}$ depends on the background field strength configuration.

As an example, consider the special case where only $F_{1\overline{1}}$ and
$F_{2\overline{2}}$ are non-zero. A gauge field configuration which reproduces
this field strength is:%
\begin{equation}
A=-F_{1\overline{1}}\overline{z_{1}}\cdot dz_{1}+F_{2\overline{2}}z_{2}\cdot
d\overline{z_{2}}\text{.}%
\end{equation}
Solving equations (\ref{one})-(\ref{three}) yields $\psi_{\overline{2}}=0$
and:
\begin{equation}
\psi_{\overline{1}}\varpropto\chi_{12}\varpropto\alpha\left(  z_{2}\right)
\cdot\exp\left(  -z_{1}\overline{z_{1}}/\sqrt{v_{2}}\right)  \cdot\exp\left(
-F_{2\overline{2}}\cdot z_{2}\overline{z_{2}}+\left(  F_{1\overline{1}}\cdot
z_{1}\overline{z_{1}}/2\right)  +O(\sqrt{v_{2}})\right)
\end{equation}
where $\alpha\left(  z_{2}\right)  $ denotes a holomorphic function of $z_{2}%
$. Adiabatically including the position dependence in the $F_{i\overline{j}}$
generates a series expansion in the $z$'s.

\bigskip

\textbf{Note added in revised version:}

\bigskip

In the context of minimal F-theory GUTs, the $5\times10\times10$ interaction
term originates from a local enhancement from $SU(5)$ to $E_{6}$. Here, the
$5$ localizes on a curve where $SU(5)$ enhances to $SU(6)$, and the $10$'s
localize on curves where $SU(5)$ enhances to $SO(10)$. A subtle point in this
construction is that when a full GUT multiplet localizes on the $SO(10)$
curve, achieving a rank one matrix requires the $10$'s to localize on the same
matter curve \cite{BHVII}. This appears to suggest the condition that the
curve on which the $10$'s localize should self-intersect, or \textquotedblleft
pinch\textquotedblright\ at the point of $E_{6}$ enhancement. However, as
pointed out in \cite{TatarCodimThree}, the wave functions of the $Q$ fields
will now have two components, $\Psi_{Q_{+}}$ and $\Psi_{Q_{-}}$, corresponding
to the profile of the $Q$ wave functions on the two local pieces of the
pinched curve. Combined with the analogous contribution from the $U$ wave
functions, this would then lead to a sum of two Yukawa couplings. Because
$\Psi_{Q_{+}}$ and $\Psi_{Q_{-}}$ a priori have different coordinate
dependence (as they localize on different pieces of the same pinched curve),
this will generically lead to a rank two matrix, unlike the rank one case
required in the analysis of this paper. A simple remedy to this issue is to
impose a $\mathbb{Z}_{2}$ symmetry which interchanges these two components of
the pinched curve. Alternatively, we can consider geometries quotiented by
this same $\mathbb{Z}_{2}$ symmetry. In the quotiented geometry, the two
enhancements from $SU(5)$ to $SO(10)$ are identified, with the $E_{6}$ point
invariant under this group action. More generally, one can consider geometries
where a $\mathbb{Z}_{2}$ action interchanges two smooth curves.

Although the existence of such $\mathbb{Z}_{2}$ quotients may appear ad hoc,
this is in fact the \textit{generic} situation in compactifications of
F-theory! As noted in \cite{BHVI}, there could be monodromies acting on the
seven-branes of F-theory. In fact such monodromies were crucial for
understanding how the non-simply laced groups arise from F-theory
compactifications \cite{FMONO}. However, such brane monodromies were not used
in the constructions of F-theory GUTs presented in \cite{BHVI}. As pointed out
in \cite{TatarCodimThree}, exactly these generic monodromies will make the $%
\mathbb{Z}
_{2}$ quotient just described above automatic!

The essential point is that in the breaking pattern $E_{6}\supset SU(5)\times
SU(2)\times U(1)$, it is a rather special condition on the nature of the
singularity to require two distinct $SO(10)$ matter curves. Indeed, typically
there will be a branch cut in the configuration, so that the two seemingly
distinct components where a $10$ localizes will interchange under monodromy
around the point of enhancement. This corresponds to a single smooth curve
with a branch cut locus emanating out from the point of enhancement.

The relevance of such geometries for phenomenology was clarified in
\cite{TatarCodimThree} where these configurations were analyzed as
deformations of an element of the Cartan of $E_{6}$ such that the
$\mathbb{Z}_{2}$ Weyl group of the $SU(2)$ factor in the above breaking
pattern permutes the two $10$'s. This follows from the fact that the $10$'s
transform as a doublet of $SU(2)$. This is the same geometric $\mathbb{Z}_{2}$
quotient discussed earlier, and shows that rather than being special, it is in
fact the more generic case from the perspective of geometry.

It is important to note, however, that just as in any orbifold theory, working
in terms of quantities invariant under the group action in the covering theory
clearly suffices in all computations. Thus, it is always enough to compute all
relevant wave function overlaps in the covering theory. Quotienting by the
appropriate Weyl group (in the $E_{6}$ breaking to $SU(5)$ case, the Weyl
group of $SU(2)$), it follows that the computation of Yukawas in the covering
theory fully determines the Yukawas in the quotient theory, and does not
modify the estimates presented in this paper.

Finally, it has also recently been found in \cite{CCHV} that background gauge field fluxes
alone do not distort the rank of the Yukawa matrix. Nevertheless, the general philosophy of this paper
that background fluxes of the compactification can alter the profile of the matter field wave
functions,  and also the structure of the Yukawa matrices was recently confirmed in \cite{CCHV} by
considering a non-commutative deformation of the F-terms of the seven-brane superpotential induced by background
$H_{(1,2)} = H_{R} + \tau_{IIB} H_{NS}$ fluxes of the compactification.
Such contributions correspond to higher dimension operators of the type
briefly alluded to in equations (\ref{gaugecond}) and (\ref{chiequation}). In some
simplified examples it was found in \cite{CCHV} that much of the hierarchical
structure of the quark and charged lepton masses, and the CKM matrix are recovered.






\begin{thebibliography}{99}
\bibitem {footnoteONE}Note that the diagonal $U(1)$ here is related to the
rotation of the canonical line bundle on $S$, with local section given by
$dz_{1}\wedge dz_{2}$.
\bibitem {axialNOTE}Indeed, the $U(1)_{1}\times U(1)_{2}$ charge of every term
obtained from expanding the above exponential is of the form $(k,-l)$ or
$(-k,l)$ for $k$ and $l$ non-negative integers. As such, it is not possible
for such terms to contribute to the Yukawa couplings, because the factors
$\left(  z_{1}\right)  ^{3-i}\left(  z_{2}\right)  ^{3-j}$ have non-negative
charges with respect to both $U(1)$'s.
\end{thebibliography}
\end{document}